# A Varying Mass-To-Light Ratio in the Galactic Center Cluster?

Prasenjit Saha [1], Geoffrey V. Bicknell [2], and Peter J. McGregor [1]

## ABSTRACT

We reanalyze published kinematic and photometric data for the cool star population in the central 10 pc ($240''$) of the Galaxy, while (a) isolating the photometric data appropriate to this population, and (b) properly allowing for projection effects. Under the assumptions that the system is spherical and isotropic, we find that $M/L_K$ varies from $\leq 1$ outside a radius of 0.8 pc to $> 2$ at 0.35 pc. This behavior cannot be due to the presence of a central massive black hole. We suggest that such a varying $M/L_K$ may be due to an increasing concentration of stellar remnants towards the Galactic center. Our derived mass-radius curve confirms the existence of $\sim 3 \times 10^6 \, M_\odot$ within 0.35 pc of the Galactic center, and $\sim 1.5 \times 10^6 \, M_\odot$ within 0.2 pc. However, the latter estimate is subject to the uncertain distribution of cool stars in this region. We also consider the dynamics of the hot star population close to the Galactic center and show that the velocity dispersion of the He I stars and the surface brightness distribution of the hot stars are consistent with the mass distribution inferred from the cool stars.

*Subject headings:* galaxies: nuclei — Galaxy: center — Galaxy: kinematics and dynamics

## 1. INTRODUCTION

It has long been conjectured that the ultimate source of energy in active galactic nuclei is a black hole with mass $\gtrsim 10^6 \, M_\odot$. This idea that many galactic nuclei harbor relic black holes continues to influence studies of our own galactic center, and is given credence by the existence of the unique radio source Sgr A* apparently at the dynamical center of the Galaxy (Backer & Sramek 1987). Sgr A* is explicable in terms of low-level accretion onto a central black hole of mass $\sim 10^6\ M_\odot$ (Melia 1994; Narayan et al. 1995). However, it is unclear whether a significantly lower mass black hole can also account for this unique object.

---

[1] Mt. Stromlo & Siding Spring Observatories, Weston PO, ACT 2611, Australia;
email: saha@mso.anu.edu.au, peter@mso.anu.edu.au

[2] ANU Astrophysical Theory Centre. Australian National University, Canberra, ACT 0200, Australia. email: geoff@wintermute.anu.edu.au. The ANUATC is operated jointly by the Mt. Stromlo and Siding Spring Observatories and the School of Mathematical Sciences.

Recent work on the stellar populations in the Galactic center (Allen, Hyland, & Hillier 1990; Krabbe et al. 1991; Krabbe et al. 1995) has shown that most of the ionizing flux, luminosity, and wind flowing from the Galactic center region arise from hot stars, with any central black hole being energetically insignificant. A similar conclusion is reached from consideration of the faintness of the potential near-infrared counterpart of Sgr A* (Herbst, Beckwith, & Shure 1993; Close, McCarthy, & Melia 1995; Eckart et al. 1995). Nevertheless, a central black hole may still be dynamically important, with studies of both gas (Serabyn et al. 1988) and stellar (McGinn et al. 1989, hereafter MSBH; Sellgren et al. 1990, hereafter SMBH) dynamics concluding that a mass in excess of $10^6\ M_\odot$ resides within the central few arcseconds. However, massive black hole models for the dynamics of our own galactic center are not without their problems. It has been argued that a $10^6\ M_\odot$ black hole would disperse the IRS 16 star cluster in $\sim 10^3$ yr (Allen & Sanders 1986), that its presence is inconsistent with the observed offset of $\sim 1''$ between IRS 16 and Sgr A*, if Sgr A* is the $10^6\ M_\odot$ black hole, and that the young luminous stars in the central region could not have formed so close to a massive black hole (Sanders 1992).

In the present paper, we reanalyze the stellar dynamics of the Galactic center region in order to determine the extent to which the presence of a massive black hole is demanded by the available data. To probe the mass distribution near the Galactic center, MSBH and SMBH measured the rotation and velocity dispersion of late-type stars within 4 pc of the Galactic center using the 2.3 $\mu$m CO absorption bandhead. Their observations revealed a remarkable feature of the velocity dispersion profile; the stellar velocity dispersion increases from $\sim 50$ km s$^{-1}$ at 4 pc ($\sim 100''$) to $\sim 120$ km s$^{-1}$ at 0.2 pc ($\sim 5''$) from the Galactic center. It is this feature of the stellar distribution which points most strongly to the presence of a massive compact object in the Galactic center. Under the simplifying assumptions that the system is spherical and the velocity dispersion is isotropic, the radial mass distribution can be obtained from stellar hydrodynamics. This was done by MSBH, who extrapolated the mass distribution to infer a central black hole mass of $2.5 \times 10^6\ M_\odot$. With additional data, SMBH concluded that the unseen mass enclosed within the central 0.6 pc ($\sim 15''$) of the Galaxy is $\sim 5 \times 10^6\ M_\odot$. Kent (1992) incorporated the MSBH and SMBH data into his more wide-ranging analysis of the Galactic bulge. Using an oblate isotropic model that included a discrete central mass, and with constant $K$ band mass-to-light ratio, he estimated the discrete central mass to be $\sim 3 \times 10^6\ M_\odot$ and the stellar $M/L_K$ to be $\sim 1$.

We were prompted to reexamine the MSBH and SMBH analysis for a number of reasons: Firstly, Allen (1994) has shown that the cool stellar population displaying 2.3 $\mu$m CO absorption has a different spatial distribution from the total light. Secondly, the analyses by MSBH and SMBH neglect the effect of projection on the velocity dispersion. This has a moderate effect on the inferred mass distribution. Thirdly, accurate mass and deprojected light distributions are necessary for the purpose of comparing enclosed mass and enclosed light, and so deriving the appropriate mass-to-light ratio. The correct interpretation of this mass-to-light ratio has an important bearing on whether one is justified in inferring the presence of a black hole.

In the following analysis, we adopt a value of 8.5 kpc for the distance to the Galactic center corresponding to a scale of 0.041 pc per arcsecond.

## 2. THE MASS MODEL

### 2.1. Outline and Assumptions

Our reanalysis of the stellar hydrodynamics of the Galactic center cluster uses observations of the radial distributions of the $K$ band surface brightness $\Sigma$, projected rotational velocity $v_{\rm p}$, and projected velocity dispersion $\sigma_{\rm p}$. To proceed, we assume that the stellar population is spherically distributed with isotropic velocity dispersion, and that it is in equilibrium in an underlying spherical gravitational potential. We show below how the potential is then determined from $\Sigma$, $v_{\rm p}$, and $\sigma_{\rm p}$, and how $M/L_K$ is obtained as a function of $r$. The problem is under constrained, and formally we could equally well have chosen to adopt other sets of assumptions. For instance, if we had assumed that $M/L_K$ remained constant we could have allowed for anisotropy in the true velocity dispersion in deriving the potential or, alternatively, have allowed for oblateness in the stellar distribution. The latter option is the one explored by Kent (1992). We note that simply relaxing any one of our assumptions without constraining another parameter leaves the potential under-determined.

The plausibility of the assumptions we have adopted was discussed by MSBH. The relaxation time of the cluster is estimated to be $\sim 10^8$ yr. This is much shorter than the age of the late-type giant population in question, so it is reasonable to assume that the system is isotropized. Furthermore, the observed $v_{\rm p}$ and $\sigma_{\rm p}$ appear to be consistent with rotational flattening, and so no large anisotropy in the velocity dispersion is likely. The ellipticity of the Galactic center star cluster is uncertain, but within $\sim 100$ pc ($\sim 40'$) of the center it is estimated to be $0.7 \pm 0.1$ (MSBH; Lindqvist, Habing, & Winnberg 1992). We show below that our assumption of sphericity does not significantly affect the derived potential.

## 2.2. Calculations

Under our assumptions, the gravitational potential is given by the spherically symmetric Jeans equation

$$\frac{1}{\rho}\frac{d}{dr}\left(\rho\sigma_{\rm r}^2\right) - \frac{v_{\rm rot}^2}{r} = -\frac{GM(r)}{r^2} \qquad (1)$$

from which $M(r)$, the total mass enclosed within radius $r$, can be inferred. Here $\rho$, $\sigma_{\rm r}$ and $v_{\rm rot}$ are the radial volume density, three-dimensional velocity dispersion, and rotational velocity functions, respectively, of a set of test particles within the potential.

The three dimensional variables are determined from the corresponding projected quantities by inverting the Abel integrals:

$$\begin{aligned}
\Sigma(r_{\rm p}) &= 2\int_{r_{\rm p}}^{\infty} \frac{\rho(r)}{\sqrt{r^2 - r_{\rm p}^2}} r\,dr, \\
\Sigma(r_{\rm p})\sigma_{\rm p}^2(r_{\rm p}) &= 2\int_{r_{\rm p}}^{\infty} \frac{\rho(r)\sigma_{\rm r}^2(r)}{\sqrt{r^2 - r_{\rm p}^2}} r\,dr, \\
\Sigma(r_{\rm p})v_{\rm p}(r_{\rm p}) &= 2\int_{r_{\rm p}}^{\infty} \frac{\rho(r)v_{\rm rot}(r)}{\sqrt{r^2 - r_{\rm p}^2}} r\,dr,
\end{aligned} \qquad (2)$$

$r_{\rm p}$ being the projected radius from the dynamical center. We carried out the deprojections using a technique and program due to A. J. Kalnajs (private communication). This first transforms equations (3) to the variables $\ln r_{\rm p}$ and $\ln r$, thus converting them into convolutions, and then solves by a Fourier transform deconvolution method. We emphasize that $\rho(r)$ is the density of test particles that are used to trace the mass distribution, whereas $M(r)$ is the total enclosed mass of stars, gas, and any dark matter such as black holes.

The use of the spherically symmetric Jeans equation is a particularly good choice when modeling data along the major axis of the Galactic center cluster for the following reasons. The axisymmetric Jeans equations in cylindrical coordinates $(R, z, \phi)$ are:

$$\begin{aligned}
\frac{\partial}{\partial R}\left(\rho\,\sigma_R^2\right) - \rho\frac{v_\phi^2}{R} &= -\rho\frac{\partial \Phi}{\partial R} \\
\frac{\partial}{\partial z}(\rho\sigma_z^2) &= -\rho\frac{\partial \Phi}{\partial z}
\end{aligned} \qquad (3)$$

where $\Phi$ is the gravitational potential and $\sigma_R = \sigma_z = \sigma_\phi$ for isotropy (Binney & Tremaine 1987). The first of the dynamical equations when restricted to the plane ($z = 0$) containing the major axis of the cluster is identical to the spherically symmetric Jeans equation (Eq. 1). Moreover, the projected variables are related to the 3-space variables by the same Abel projection integrals since the geometry of the projection is identical. Thus, in effect, in using equation (1) we are estimating $-\partial\Phi/\partial R$ and equating this to $-GM(r)/r^2$. The mass profile so obtained should be reasonably indicative of the mass profile of the slightly oblate Galactic center region, and especially so since the gravitational potential is more spherical than the test particle density in the region that produces it.

An Abel projection is in effect a convolution so that the reverse process of deprojection is subject to all of the problems associated with deconvolution. In particular, if the data are noisy the deprojection is unreliable. Hence, we model the surface brightness, projected rotational velocity, and projected velocity dispersion using functional forms. For the surface brightness, we use the Reynolds-Hubble law often used for fitting extragalactic surface brightness profiles. Thus we assume that

$$\Sigma(r_{\rm p}) = \Sigma_0 \left(1 + \frac{r_{\rm p}^2}{r_{\rm p,0}^2}\right)^{-\alpha}. \tag{4}$$

One feature of this expression is that it can be deprojected exactly, to give:

$$\rho(r) = \frac{\Sigma_0}{r_{\rm p,0} B(\frac{1}{2}, \alpha)} \left(1 + \frac{r^2}{r_{\rm p,0}^2}\right)^{-(\alpha+\frac{1}{2})}. \tag{5}$$

where $B$ is the Beta function. We fit the projected rotation velocity and velocity dispersion, $v_{\rm p}$ and $\sigma_{\rm p}$, with quartics in the variable $u = \ln(r_{\rm p,0} + r_{\rm p})$ such that

$$\sigma_{\rm p}(r_{\rm p}) = a_0 + a_1 u + a_2 u^2 + a_3 u^3 + a_4 u^4, \tag{6}$$
$$v_{\rm p}(r_{\rm p}) = b_0 + b_1 u + b_2 u^2 + b_3 u^3 + b_4 u^4. \tag{7}$$

A logarithmic variable is mathematically convenient for fitting data over a wide range in radius, and using $\ln(r_{\rm p,0} + r_{\rm p})$ avoids a singularity in the fit at $r_{\rm p} = 0$. By fitting the observational data with smooth functions, we eliminate spurious local gradients which would otherwise be amplified by deprojection.

### 2.3. Observational Data

The reliability of our mass estimates depends on the reliability of the surface brightness, projected velocity dispersion, and projected rotational velocity data for a well-defined set of test particles in the Galactic center cluster. Our understanding of this region has improved recently with the realisation that the central $\sim 1$ pc ($\sim 24''$) is populated by two distinct stellar populations. One is a continuation of the Galactic bulge population to smaller radii. The $K$ light from this population is dominated by red giant stars that exhibit

2.3 $\mu$m CO absorption bands in their $K$ band spectra. The other is a population of massive ($M \geq 20\ M_\odot$), evolved, hot stars that have been identified by the strong He I 2.058 $\mu$m line emission produced in their mass loss winds (Allen, Hyland, & Hillier 1990; Krabbe et al. 1991). The most prominent of these stars appear to be Ofpe/WN9 stars, a rare class of extremely massive stars in an evolutionary stage on the way to becoming Wolf-Rayet stars (McGregor, Hillier, & Hyland 1988). If these stars formed in a starburst $\sim 10^7$ yr ago, they should be associated with a larger population of slightly lower mass stars still on or near the main sequence. Such stars would be more than one order of magnitude fainter than the evolved He I emission-line stars at 2 $\mu$m, and until recently it has proved difficult to identify these stars observationally.

Burton & Allen (1992) have used the presence or absence of the 2.3 $\mu$m CO absorption band in Galactic center cluster stars to distinguish between "cool" stars and "hot" stars. Allen (1994) found that the "hot" population is confined to within a radius of $\sim 1$ pc ($\sim 24''$) of the Galactic center, with a core radius of $\leq 0.2$ pc ($\sim 5''$). In contrast, the "cool" population is more extended with a core radius of $\sim 0.6$ pc ($\sim 15''$). Similar conclusions have also been reached by Krabbe et al. (1991) from the spatial distribution of He I emission-line stars and Rieke & Lebofsky (1987) from consideration of the diffuse background light between bright stars in the central region. Since the most extensive sets of available velocity dispersion data are based on measurements of the diffuse light CO absorption bandheads (MSBH; SMBH), we adopt the "cool" star population as test particles for probing the gravitational potential, and use the $K$ band surface brightness distribution for the "cool" star population, as defined by Allen (1994), in our analysis.

SMBH have shown that the diffuse light CO absorption band weakens inside of $\sim 0.6$ pc ($15''$) of the center. They infer from this that the CO absorbing stars projected on the central region may be foreground and background objects at larger true distance, and so their velocity dispersion may not reflect the true velocity dispersion inside the central region. Haller et al. (1996) confirm the effect using a more extensive dataset which shows that the weakening occurs only inside a projected radius of $0.35 \pm 0.06$ pc ($8.5 \pm 1.5''$). Hence, mass estimates based on the "cool" star population interior to $\sim 0.35$ pc should be treated with some caution due to the uncertain density distribution of these stars.

Figure 1 shows the $K$ band surface brightness data for the "cool" star and "hot" star populations referred to the position of IRS 16 (Allen 1994). Although IRS 16 is offset by $\sim 1''$ from Sgr A$^*$, the likely dynamical center (Backer & Sramek 1987; Eckart et al. 1993, 1995), this distinction is not significant for our analysis because the innermost data point considered is at a radius of $4.4''$ (see below). The contribution from the M supergiant, IRS 7, has been removed from the "cool" star data. In fitting the surface brightness distribution, we arbitarily adopt errors of ±0.1 mag for these measurements. Projected surface brightness data from Becklin & Neugebauer (1968; 1975), as compiled by Bailey (1980), can be fit to a radius of at least 30 pc using a power-law with index of -0.8. We therefore constrain our fit to asymptote to this slope at large radii by setting $\alpha = 0.4$ in equation (4). The Bailey (1980) data points are plotted in Figure 1, but were not used in our fit. It is not necessary to convert the observed $K$ band surface brightness distribution to an absolute scale when calculating the enclosed mass because the scaling factor for density divides out of equation (1). However, it is necessary to do this when computing the mass-to-light ratio. The units of the surface brightness data were converted from mag arcsec$^{-2}$, $\mu_{K,\mathrm{obs}}$, to $K$ band solar luminosity per square parsec, $\Sigma$, using

$$\log \Sigma = 2.77 - 0.4 \left( \mu_{K,\mathrm{obs}} - A_K - (m-M) - M_{K_\odot} \right). \tag{8}$$

Here, the constant converts square arcsecs to square parsecs, the $K$ band extinction, $A_K$, is taken to be 3.5 mag (Rieke, Rieke, & Paul 1989), the distance modulus, $m - M$, is 14.65 for our adopted distance to the Galactic center of 8.5 kpc, and the absolute $K$ band magnitude of the Sun is determined from its absolute $V$ magnitude (Allen 1973) and the $V - K$ color appropriate for a G2V star (Johnson 1966) which leads to $M_{K_\odot} = 3.39$. Our fit to the "cool" star surface brightness data is then given by the Reynolds-Hubble profile of equation (4) with $r_{\mathrm{p},0} = 0.34$ pc and $\Sigma_0 = 3.8 \times 10^6 \, L_\odot \, \mathrm{pc}^{-2}$.

The total light profile is used in computing the mass-to-light ratio. We obtain this by separately fitting the Allen (1994) "hot" star data with $r_{\mathrm{p},0} = 0.06$ pc, $\Sigma_0 = 1.2 \times 10^7 \, L_\odot \, \mathrm{pc}^{-2}$, and $\alpha = 0.66$ in equation (4) and combining both profiles.

From recent observations, Krabbe et al. (personal communication) argue that the cool star population is better fit using a core radius $\leq 5''$ (three times smaller than implied by our fit) plus a central minimum. We have therefore obtained a second fit by reducing the parameter $r_{\mathrm{p},0}$ by a factor of 3 (to 0.11 pc), again constrained $\alpha$ to 0.4, and fit the Allen (1994) data with $\Sigma_0 = 7.3 \times 10^6 \, L_\odot \, \mathrm{pc}^{-2}$ (Fig. 1, dashed line). Note that this fit is not a good one inside the innermost velocity dispersion point at 0.18 pc (4.4''), but this is of no consequence. We show below that this reduction in core radius makes no significant difference to the inferred mass profile outside 0.2 pc.

Figures 2 and 3 show our fits to the projected velocity dispersion and rotational velocity data from MSBH and SMBH, and Group I OH/IR star kinematic data from Lindqvist, Habing, & Winnberg (1992), again referred to IRS 16 by these authors. The quartic coefficients for these fits are listed in Table 1. Figures 2 and 3 also plot the deprojected quantities, $\sigma_{\mathrm{r}}(r)$ and $v_{\mathrm{rot}}(r)$, that enter into equation (1). The OH/IR stars are mostly well outside the central few parsecs; we include them in order to better constrain the behavior of our kinematic fits at large projected radius, but our results are not very sensitive to this.

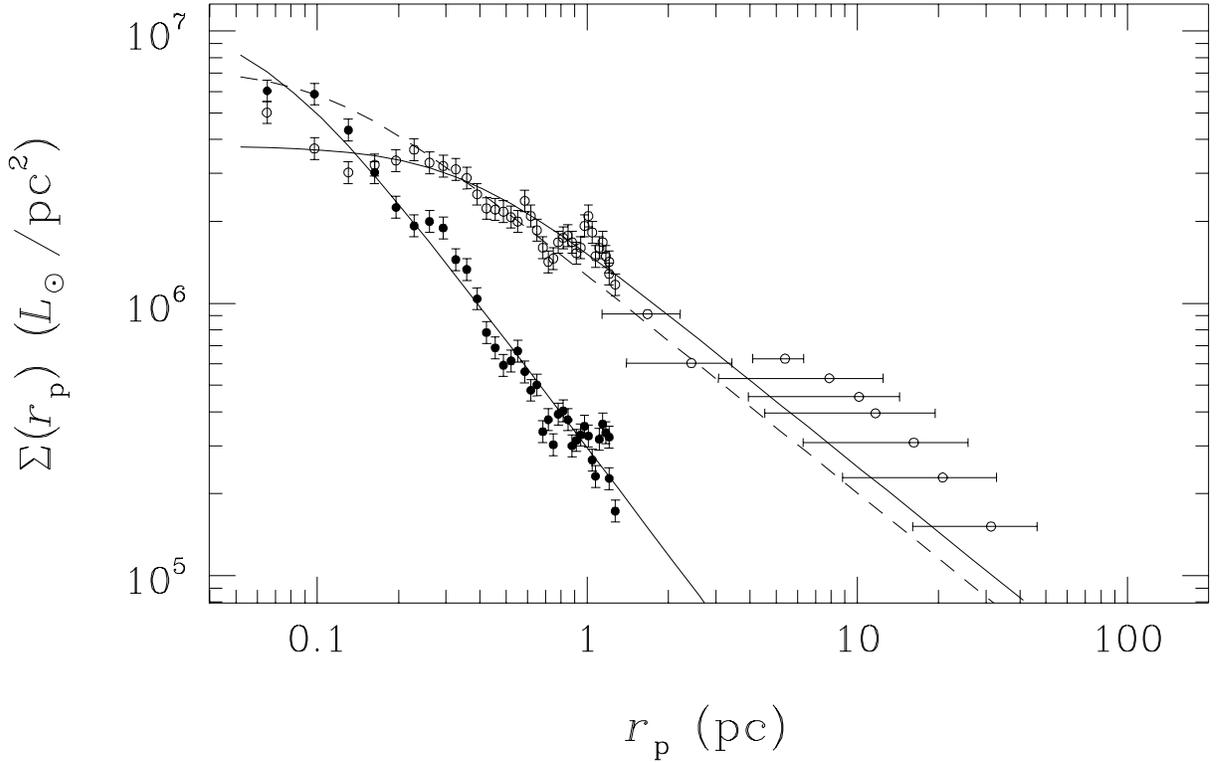

Fig. 1.— Surface brightness fits as a function of projected radius (solid lines). The data are from Allen (1994; vertical error bars), and Becklin & Neugebauer (1968; 1975) as compiled by Bailey (1980; horizontal error bars). Data for "cool" stars with 2.3 $\mu$m CO absorption bands (open symbols) and "hot" stars lacking 2.3 $\mu$m CO absorption bands (solid symbols) are shown separately. A fit to the "cool" star data with the core radius reduced by a factor of three is also shown (dashed line; see text)

We have used only the Group I OH/IR stars of Linqvist et al. because these are thought to belong to the same dynamical population as the cool stars making up the Galactic center cluster. Our OH/IR star velocity dispersion data are dispersions about the mean velocity of groups of $\sim$ 20 stars ordered in increasing galactic longitude, and do not depend on the Lindqvist et al. linear fit to galactic rotation.

Table 1. Velocity Fit Coefficients

| $\sigma_{\rm p}$ | Fit 1 | Fit 2 | $v_{\rm p}$ | Fit |
|---|---|---|---|---|
| $a_0$ | 113.49 | 98.20 | $b_0$ | 38.38 |
| $a_1$ | -39.98 | -45.71 | $b_1$ | 14.75 |
| $a_2$ | 4.507 | 17.39 | $b_2$ | -6.878 |
| $a_3$ | 2.021 | -1.28 | $b_3$ | 0.875 |
| $a_4$ | -0.215 | 0.001 | $b_4$ | -0.023 |

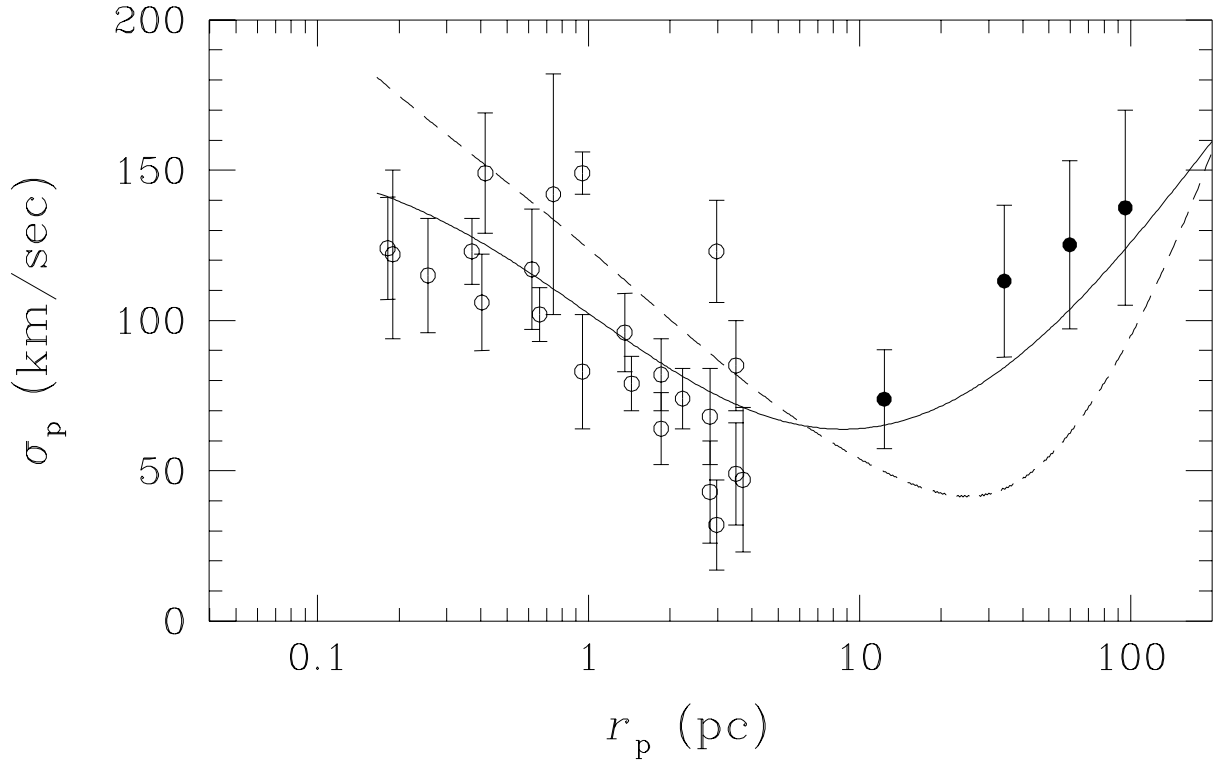

Fig. 2.— Stellar velocity dispersions for cool stars in the Galactic center region. The data within 4 pc are from McGinn et al. (1989) and Sellgren et al. (1990), omitting points thought by those authors to be contaminated by foreground stars. The four points between 10 and 100 pc are derived from discrete velocities of Group I OH/IR stars in Lindqvist et al. (1992). The solid curve shows our fit to the data, while the dashed curve is its deprojection, $\sigma_r$.

Several points in Figure 2 lie above the bulk of the MSBH and SMBH data points and may be the result of observational errors or local velocity dispersion anomalies. Our fit to the velocity dispersion data has been repeated with four apparently anomalous data points omitted in order to gauge their effect. This fit is shown in Figure 4 and the quartic coefficients are listed in Table 1.

### 2.4. Distribution of Enclosed Mass

Figures 5–7 show the radial distribution of enclosed mass derived from our numerical solution of equation (1) and the separate fits to the surface brightness and velocity dispersion profiles. In each figure, the dashed curves represent one sigma deviations of the

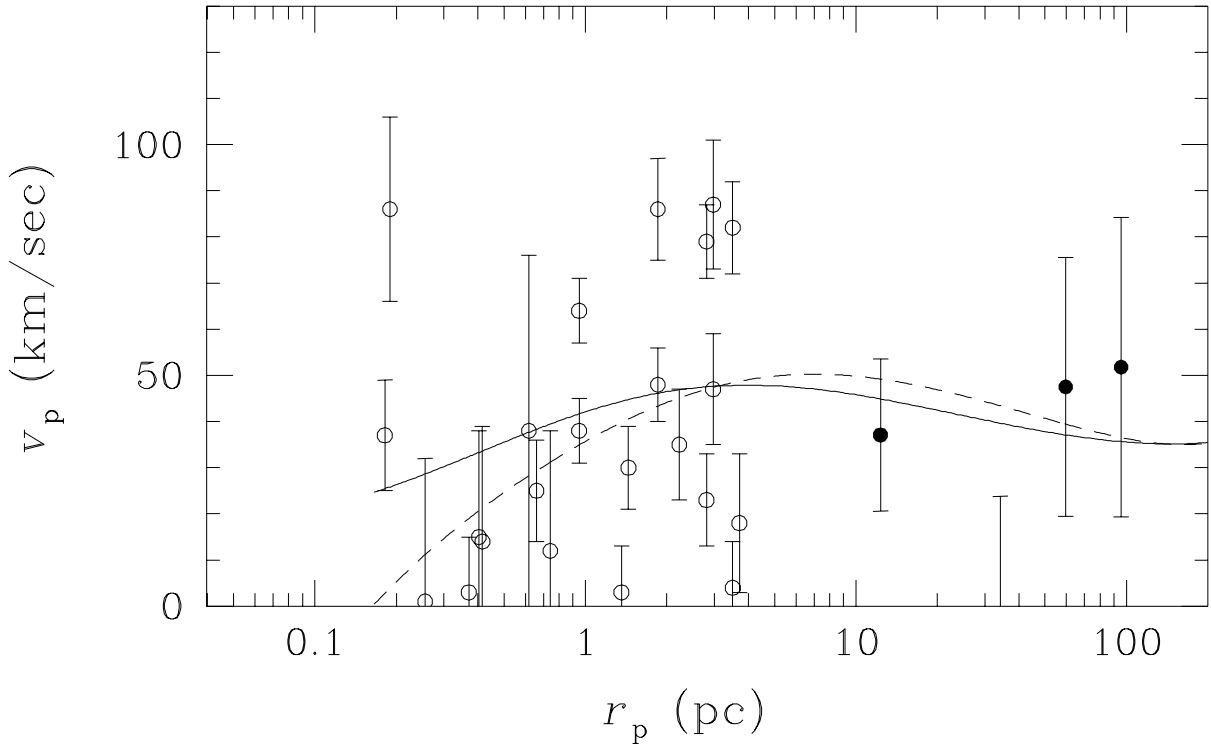

Fig. 3.— Similar to Fig. 2, but for rotation. The solid curve shows our fit to the data, and the dashed curve represents deprojected rotation.

estimated mass. These were derived by first generating 20 fictitious kinematic data sets by adding noise in accordance with the published error estimates to the input data. Then the dynamical analysis was performed on each fictitious data set to obtain an ensemble of 20 mass profiles. Our error estimate on $M(r)$ is the dispersion in this ensemble at each radius, $r$. Also plotted in Figures 5–7 are the enclosed mass estimates of MSBH and the mass estimates of Güsten et al. (1987) based on the rotational velocity of the HCN molecular ring and corrected to our adopted Galactic center distance of 8.5 kpc.

Comparison of Figures 5 and 6, corresponding to the different fits to the radial surface brightness profile, shows that the smaller core radius has minimal effect on the inferred mass profile. This confirms that our derived masses are not significantly influenced by the precise form of the surface brightness distribution in the central region.

On the other hand, the effect of omitting the four apparently anomalous velocity dispersion points is to significantly lower the derived mass. The mass-radius curve based on this fit is in better agreement with the MSBH points, although our analysis procedure is more refined. More significantly, this fit is also in better agreement with the mass inferred from HCN observations of the rotational of the molecular ring (Güsten et al. 1987), giving some justification that this fit may be preferable. However, we note in §2.5 that the value of $M/L_K$ implied by this fit may be too low.

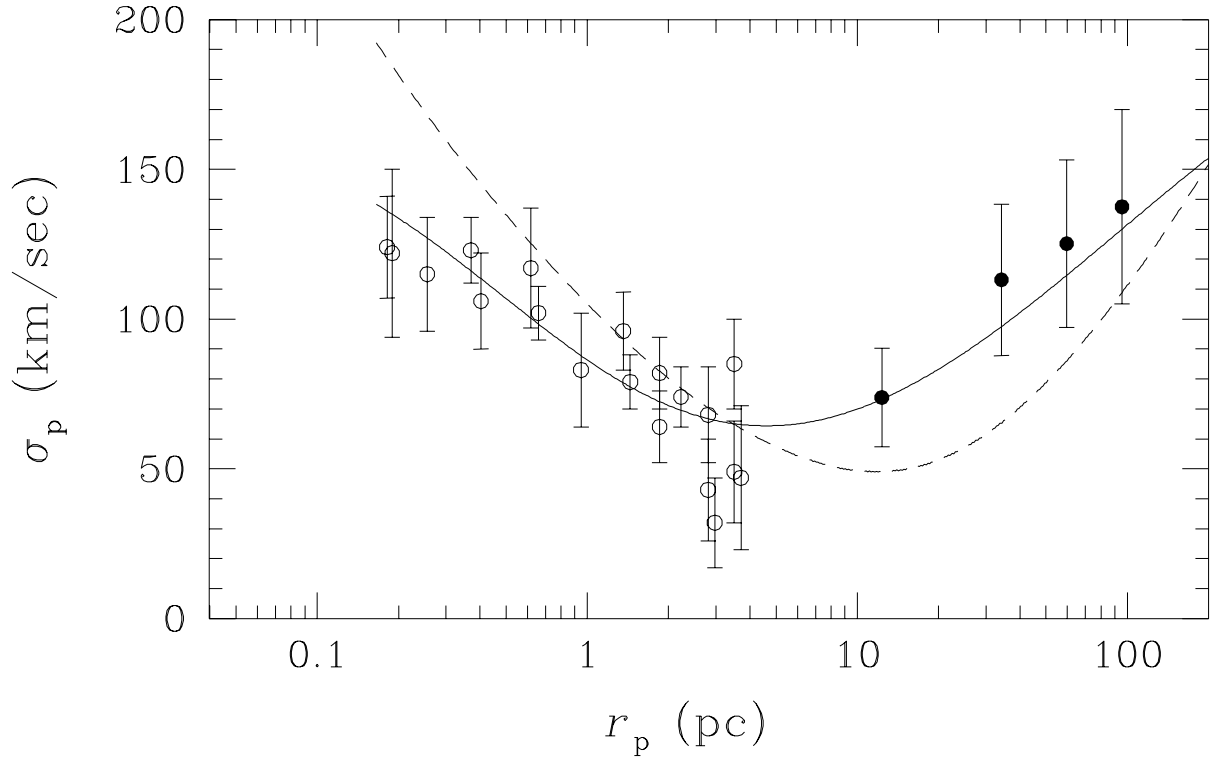

Fig. 4.— Fits to the stellar velocity dispersions for cool stars in the Galactic center region as in Fig. 2 except four outlying data points have been omitted from the fit.

The enclosed mass estimates in Figures 5 and 7 for the two velocity dispersion fits are $2.7 \times 10^6 \ M_\odot$ and $3.1 \times 10^6 \ M_\odot$, respectively, within 0.35 pc where the CO absorption band begins to weaken, and $1.2 \times 10^6 \ M_\odot$ and $1.7 \times 10^6 \ M_\odot$, respectively, inside 0.2 pc corresponding to the innermost velocity dispersion point. We repeat the above caveat that estimates interior to 0.35 pc may be subject to the uncertain spatial distribution of CO absorption stars in the central region.

## 2.5. Mass-To-Light Ratio

Additional insight into the nature of our solutions for the mass profile can be gained by plotting enclosed mass versus total enclosed $K$ band luminosity resulting from the two velocity dispersion fits (Figs. 8 and 9). The enclosed light is obtained in a consistent way from our analysis by radially integrating the deprojected surface brightness data for both the "cool" and "hot" star populations (Allen 1994). In the simple situation of a central black hole of mass $M_{\rm bh}$ and luminosity $L_{\rm bh}$ immersed in a stellar distribution with constant mass-to-light ratio $A$, it is straightforward to show that the total mass, $M$, and luminosity, $L$, within a radius, $r$, are related by

$$M(r) = (M_{\rm bh} - AL_{\rm bh}) + AL(r) \simeq M_{\rm bh} + AL(r). \qquad (9)$$

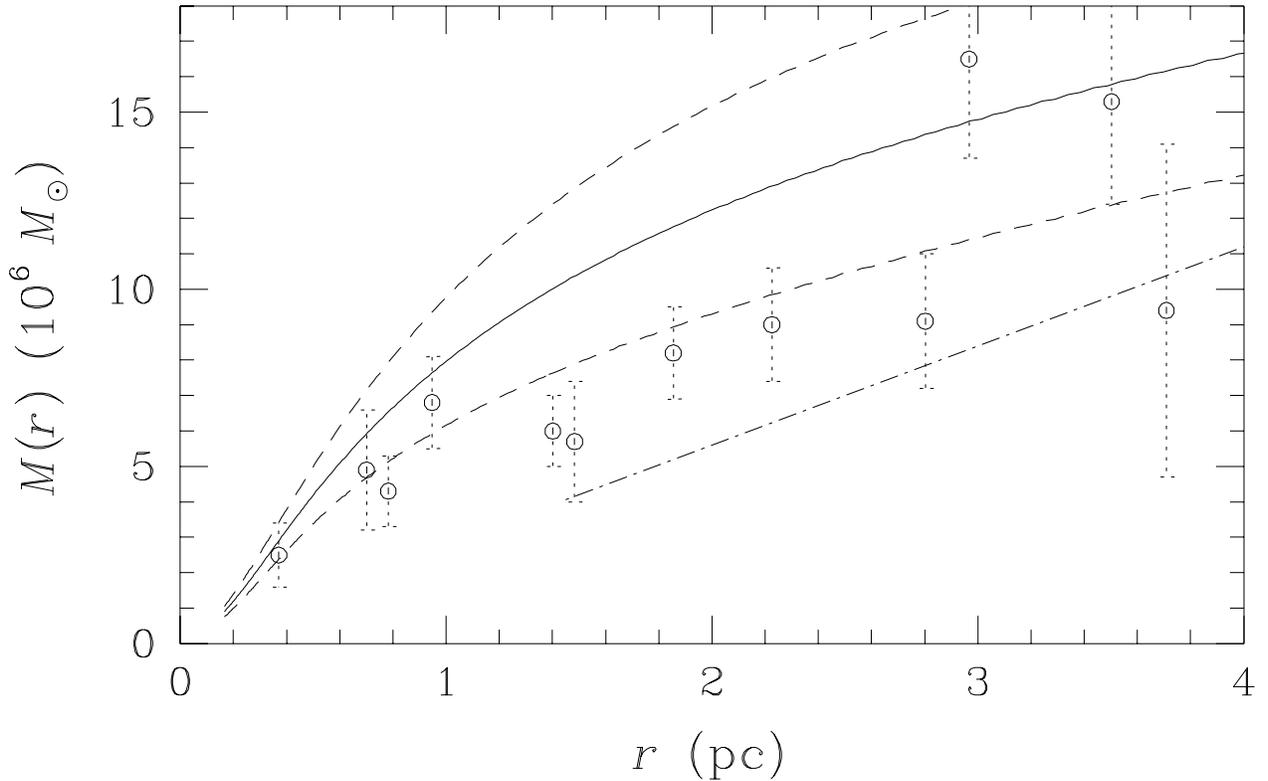

Fig. 5.— Enclosed mass as a function of radius. The solid curve shows the enclosed mass derived from our model using the first velocity dispersion fit, and the dashed curves are $1\sigma$ error estimates from Monte-Carlo variation of the kinematic data (see text). The points with dashed error bars are enclosed mass estimates from McGinn et al. (1989). The enclosed mass estimates from Güsten et al. (1987), corrected to a distance of 8.5 kpc, are indicated by a dot-dashed line.

Therefore, except in the unlikely case that the mass-to-light ratio of the black hole is identical to that of the stars, a plot of $M$ against $L$ would take the form of a straight line with slope $A$ and non-zero intercept at $\sim M_{\rm bh}$. This is not what is seen in Figures 8 and 9; the enclosed mass versus enclosed $K$ band luminosity plots show curves that gradually steepen towards decreasing radius and project to the vicinity of the origin. Within the limitations of the assumptions of our dynamical model, we are forced to the conclusion that the local $K$ band mass-to-light ratio, $M/L_K$, as determined by the local slope of the enclosed mass versus enclosed $K$ band luminosity curve, increases significantly within a radius of $\sim 0.8$ pc. The slope of the $M - L$ curve resulting from the first fit to the velocity dispersion data decreases from $\sim 0.4$ at $r = 2$ pc to $\sim 2.6$ at $r = 0.35$ pc. Similarly, the second $M - L$ curve, based on the lower velocity dispersion fit, shows a change in mass-to-light ratio from $\sim 0.22$ to 2.9 at the same radii. A similar result follows from the MSBH analysis (McGinn, private communication). The presence of a central, massive black hole in the Galactic center cannot cause this effect. In essence, our model requires more mass in regions adjacent to the Galactic center than is naively suggested by the light distribution in order to account for the rising velocity dispersion profile (Fig. 2).

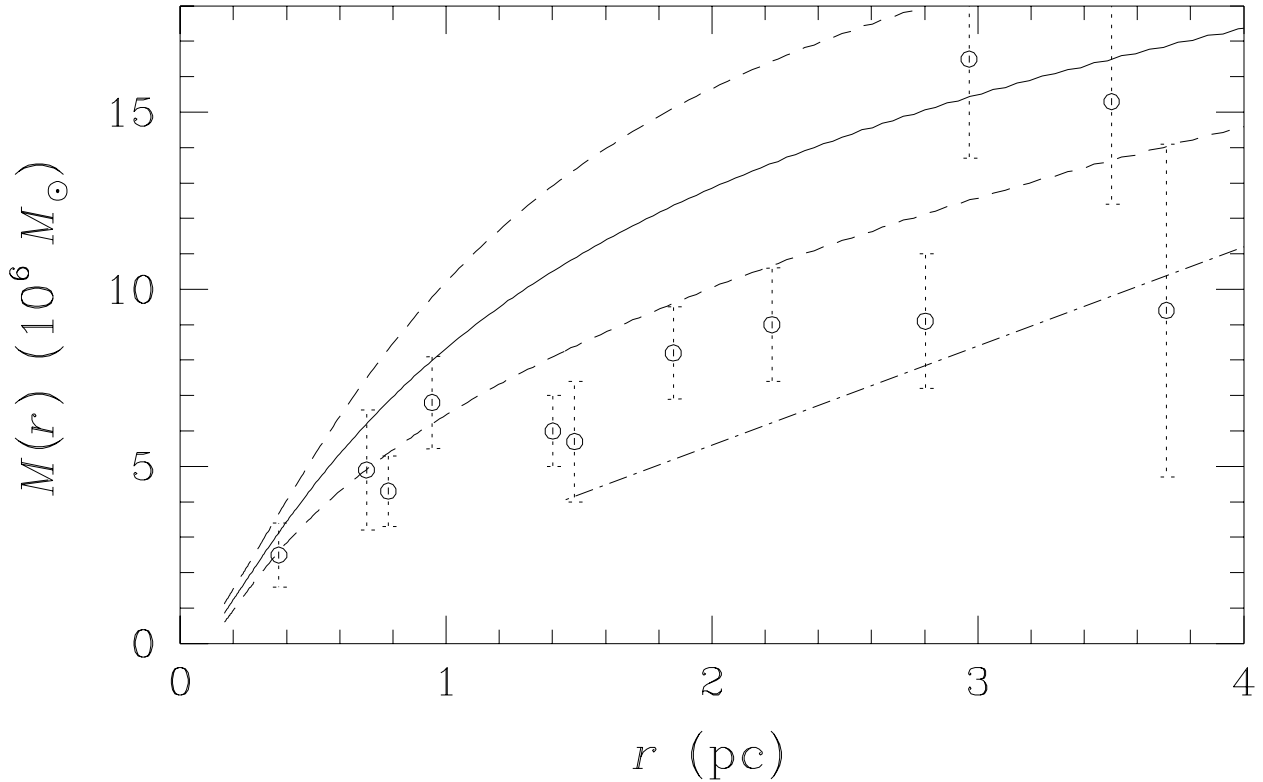

Fig. 6.— Enclosed mass as a function of radius as in Figure 5, except the mass estimates are based on the compact "cool" star surface brightness distribution fit shown as a dashed line in Fig. 1.

The inferred $M/L_K$ values at $r \gtrsim 1$ pc are smaller than the value of $\sim 1$ inferred by Kent (1992) for the Galactic bulge. This may be due to the relatively high reddening value ($A_K = 3.5$ mag) we have adopted; lowering the reddening to $A_K = 2.7$ mag (Becklin et al. 1978) increases $M/L_K$ by a factor of 2.1. Notwithstanding such a correction, the asymptotic mass-to-light ratio of 0.46, resulting from the second fit to the velocity dispersion data, may still be too low for an evolved stellar population. This issue can only be settled using more accurate velocity dispersion data for $0.5 < r < 2$ pc.

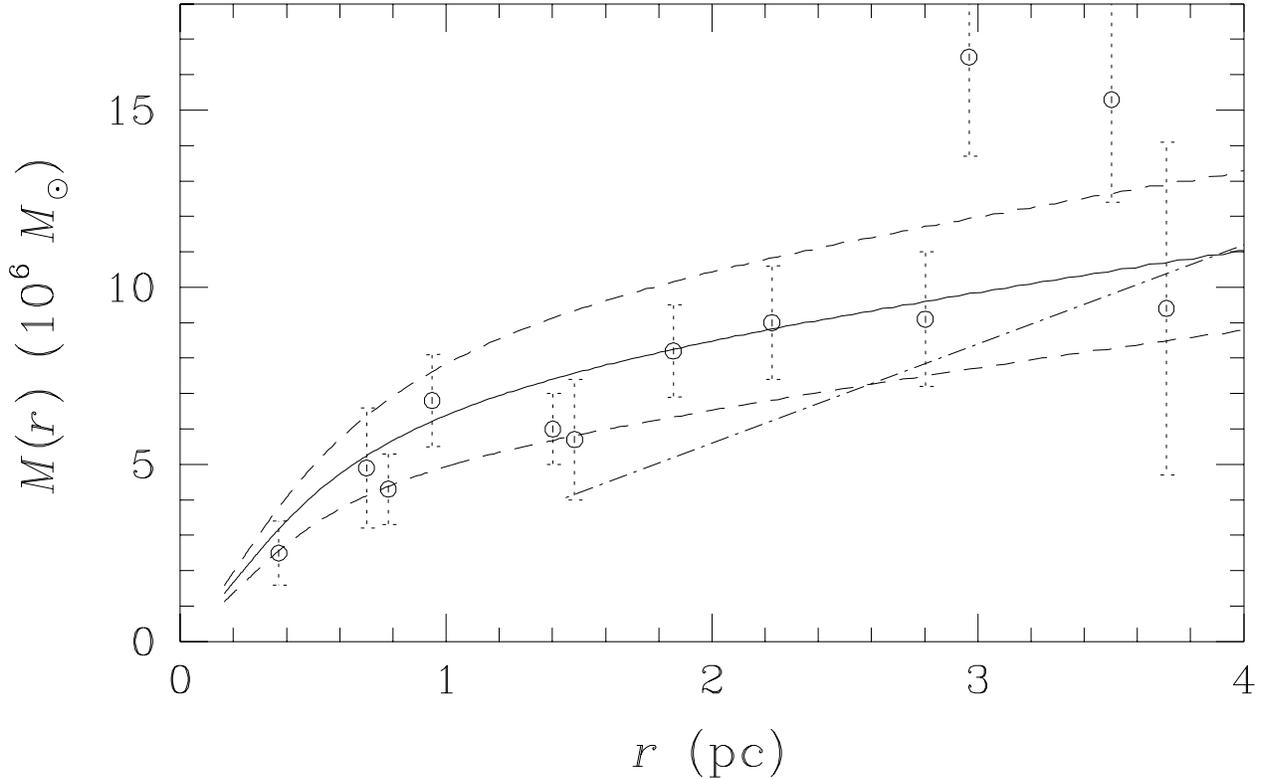

Fig. 7.— Enclosed mass as a function of radius as in Figure 5, except the mass estimates are based on the fit to the velocity dispersion data in Figure 4 that omits four outlying data points.

Given the inferred masses and asymptotic mass-to-light ratios, the two mass solutions imply $6.1 \times 10^6\ M_\odot$ and $5.3 \times 10^6\ M_\odot$ of dark mass, respectively, within 1 pc. This gives ratios of dark-to-bright mass within this region of 3.3 and 5.2, respectively. These numbers are used below in a simple model for the stellar evolution of the Galactic center cluster.

MSBH and SMBH both discussed the possibility that $M/L_K$ increases with decreasing radius, but did not quantify the range in radius or the value of $M/L_K$ required.

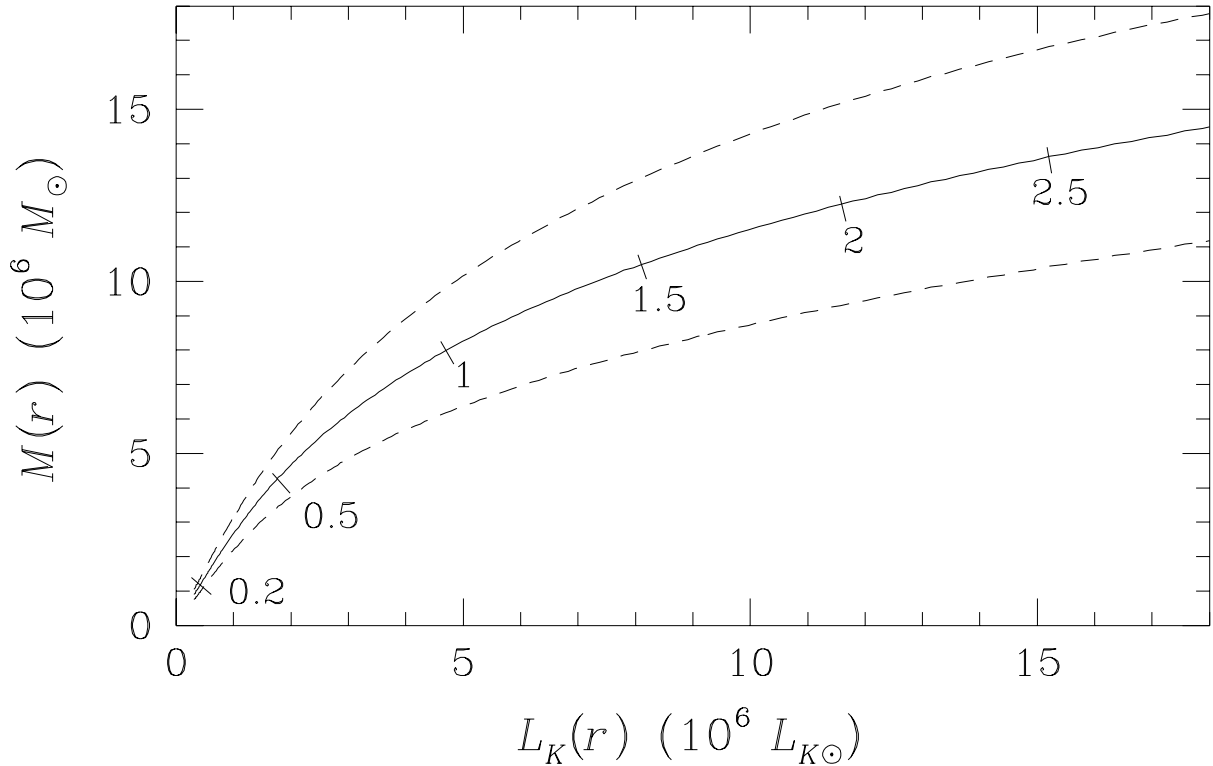

Fig. 8.— Enclosed mass at given radius versus total enclosed $K$ band luminosity in the same volume based on the first velocity dispersion fit. The enclosed luminosity is the sum of the "cool" star and "hot" star distributions. Tick marks along the curve indicate the corresponding radii in parsecs.

## 3. DISCUSSION

### 3.1. A Varying Mass-To-Light Ratio?

We now discuss some possible alternative explanations to a varying $M/L_K$ in the vicinity of the Galactic center. Anisotropy in the velocity dispersion can mimic changes in $M/L$. However, as MSBH have argued (see §2), the velocities appear to be isotropic based upon the relatively short relaxation time and the consistency of the oblateness with isotropic, rotationally flattened models. Therefore we have not considered such a possibility here. It should be noted, however, that mass segregation effects may be important in an isotropic system.

It is possible that M supergiants in the central region could affect the "cool" star $K$ band surface brightness distribution without contributing to the diffuse light CO bandhead velocity dispersion, even though the bright M supergiant, IRS 7, has been removed from the Allen (1994) "cool" star data. We gauged the effect of these stars by arbitrarily decreasing the "cool" star central surface brightness by a factor of two, and correspondingly increasing the core radius so that the surface brightness at $r_p > 1$ pc remained unchanged. The lower test particle density in the central region resulted in a higher deprojected velocity dispersion gradient in this region. This led to a 50% increase in the enclosed mass within 0.2 pc, and a decrease in the local $M/L_K$ at this radius from $\sim 2.6$ to $\sim 2.1$. A factor of two reduction in the central surface brightness greatly overestimates the contribution of M supergiants to the "cool" star light, so their presence in the central region cannot have a significant effect on the behavior of $M/L_K$ predicted by our model. Similarly, the inferred $M/L_K$ variation cannot be produced by any reasonable variation of interstellar extinction with radius.

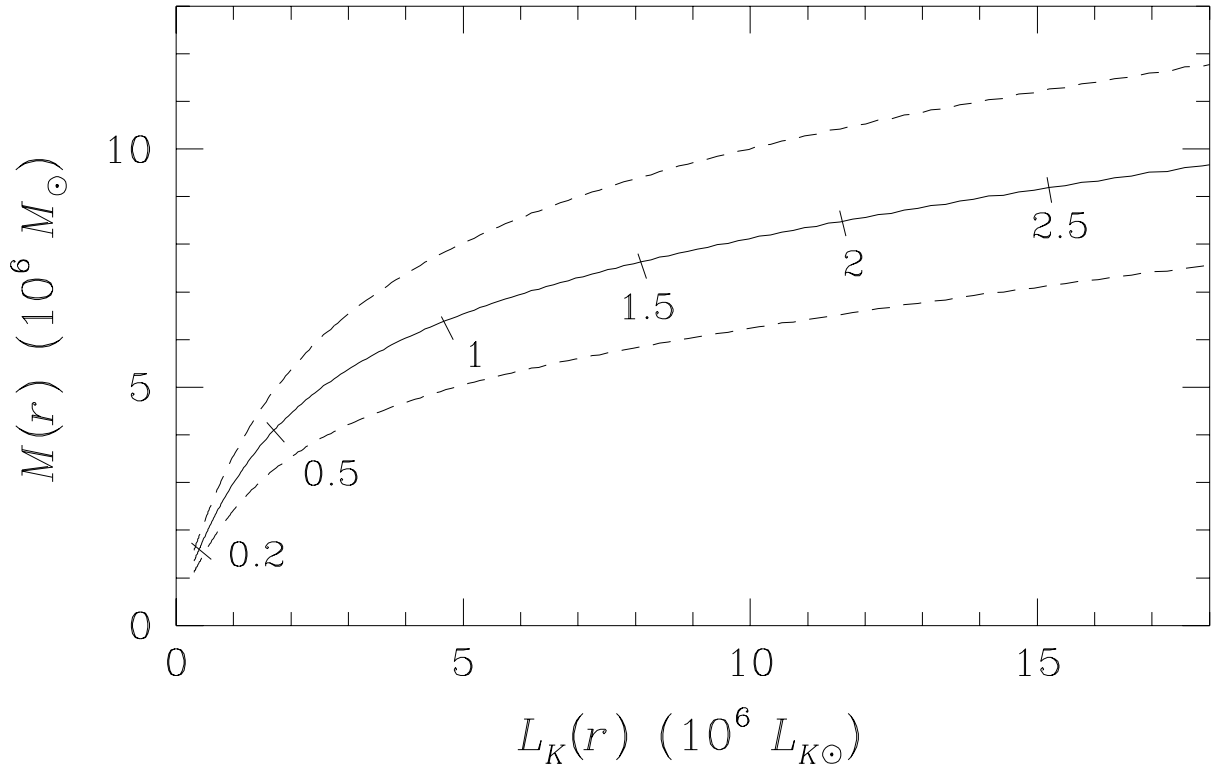

Fig. 9.— Enclosed mass at given radius versus total enclosed $K$ band luminosity in the same volume as in Figure 8, except the mass estimates are from Figure 7 and are based on the second fit to the velocity dispersion data in Figure 4 that omits four outlying data points.

Several dust enshrouded stars are present in the central region, and these stars are undoubtedly under-luminous in the $K$ band due to their high individual extinctions (Rieke, Rieke, & Paul 1989). However, these are very luminous objects and are expected to be rare, so it is unlikely that a significant stellar mass is hidden from view at 2 $\mu$m in this way. Similarly, no significant gas mass is present in the Galactic center; the mass of neutral material interior to the molecular ring is $\sim 300\ M_\odot$ (Jackson et al. 1993), and the mass of ionized material in the central region has been estimated from the radio free-free flux density to be $\lesssim 60\ M_\odot$ (Lo & Claussen 1983).

We conclude that our analysis seriously raises the prospect that $M/L_K$ varies with radius due to an intrinsic variation in the nature of the distributed stellar population in the vicinity of the Galactic center. The region of increasing $M/L_K$ coincides spatially with the central cluster of young, massive, He I emission-line stars (Krabbe et al. 1991). These stars cannot by themselves be responsible for the increasing $M/L_K$ because $M/L_K < 1$ for the luminous material in a starburst if the initial mass function is normal (Buzzoni 1989). However, such estimates do not include the mass of material from previous generations of stars that is now locked up in stellar remnants. Two possibilities involving stellar remnants can be considered.

Morris (1993) has discussed the consequences of repeated, widespread, massive starbursts in the Galactic center region. If such starbursts have occurred, he expects that over time the $\sim 10\ M_\odot$ black hole remnants of these massive stars will congregate near the Galactic center due to the effects of dynamical friction. Morris estimates that stellar remnants with masses $\sim 10\ M_\odot$ will be drawn in from radii of up to $\sim 4$ pc over $10^{10}$ yr, and that the total mass of remnants within the inner few tenths of a parsec could be as high as $0.4$–$5 \times 10^6\ M_\odot$ possibly accounting for the unseen mass of $\sim 6 \times 10^6\ M_\odot$ within 1 pc. Nevertheless, this explanation does require that large numbers of $\sim 10\ M_\odot$ black holes have formed in the region surrounding the Galactic center.

A less extreme alternative is to postulate that the bulk of the unseen mass is comprised of neutron stars and white dwarfs formed as a direct result of periodic starbursts in the central 2 pc, similar to the starburst that occurred in the Galactic center $\sim 10^7$ yr ago. We have estimated the significance of such remnants using the Salpeter (1955) initial mass function with an upper mass limit of 120 $M_\odot$ and lower mass limits of 10, 5, 2, and 1 $M_\odot$. A star formation rate of $10^{-3}\ M_\odot\ \mathrm{yr}^{-1}$ was used in order to form a total stellar mass of $10^7\ M_\odot$ in $10^{10}$ yr. Stellar lifetimes, $t_*$, from Maeder & Meynet (1988) were parameterized as a function of stellar mass, $m_*$, by

$$t_* = 12.6(m_*/M_\odot)^{-2.7} \quad \mathrm{Gyr}. \qquad (10)$$

This will over-estimate the dark mass formed in the last $\sim 10^7$ yr, but gives the correct result on longer timescales. We follow Haller et al. (1996) in assigning a black hole mass of 10 $M_\odot$ to stars with $m_* \geq 25\ M_\odot$, a neutron star mass of 1.5 $M_\odot$ to stars with $3 < (m_*/M_\odot) < 25$, and a white dwarf mass of 0.7 $M_\odot$ to stars with $1 < (m_*/M_\odot) \leq 3$. The resulting luminous stellar and dark remnant masses are shown in Figure 10(a) as functions of the population age. Truncated initial mass functions lead to a constant luminous star mass once the oldest stars of lowest mass form remnants. At the same time, the total remnant mass rises monotonically. The corresponding ratio of dark-to-bright mass, $M_{dark}/M_{bright}$, is shown in Figure 10(b). If, as seems reasonable, star formation has been continuing in the Galactic center region for $\sim 10^{10}$ yr, the $M_{dark}/M_{bright}$ ratio of $\sim$ 3–5 required by our dynamical model (see §2.5) can only be obtained if the initial mass function in the center of the Galaxy is truncated at $\sim$ 1–2 $M_\odot$. Shorter star forming durations require initial mass functions truncated at even higher masses to suppress the integrated luminosity of low mass stars. Morris (1993) has noted that the extreme conditions in the Galactic center region are likely to lead to an initial mass function favoring the formation of high mass stars. In this context, we note the agreement between the radius at which $M/L_K$ begins to increase and the extent of the "hot" star distribution of $\sim$ 1 pc (Allen 1994). Both may be set by the size of the region where star formation is able to proceed. Sanders (1992) has suggested that this is limited by the core radius of the "cool" star distribution since gravitational shear will disrupt molecular clouds beyond this radius.

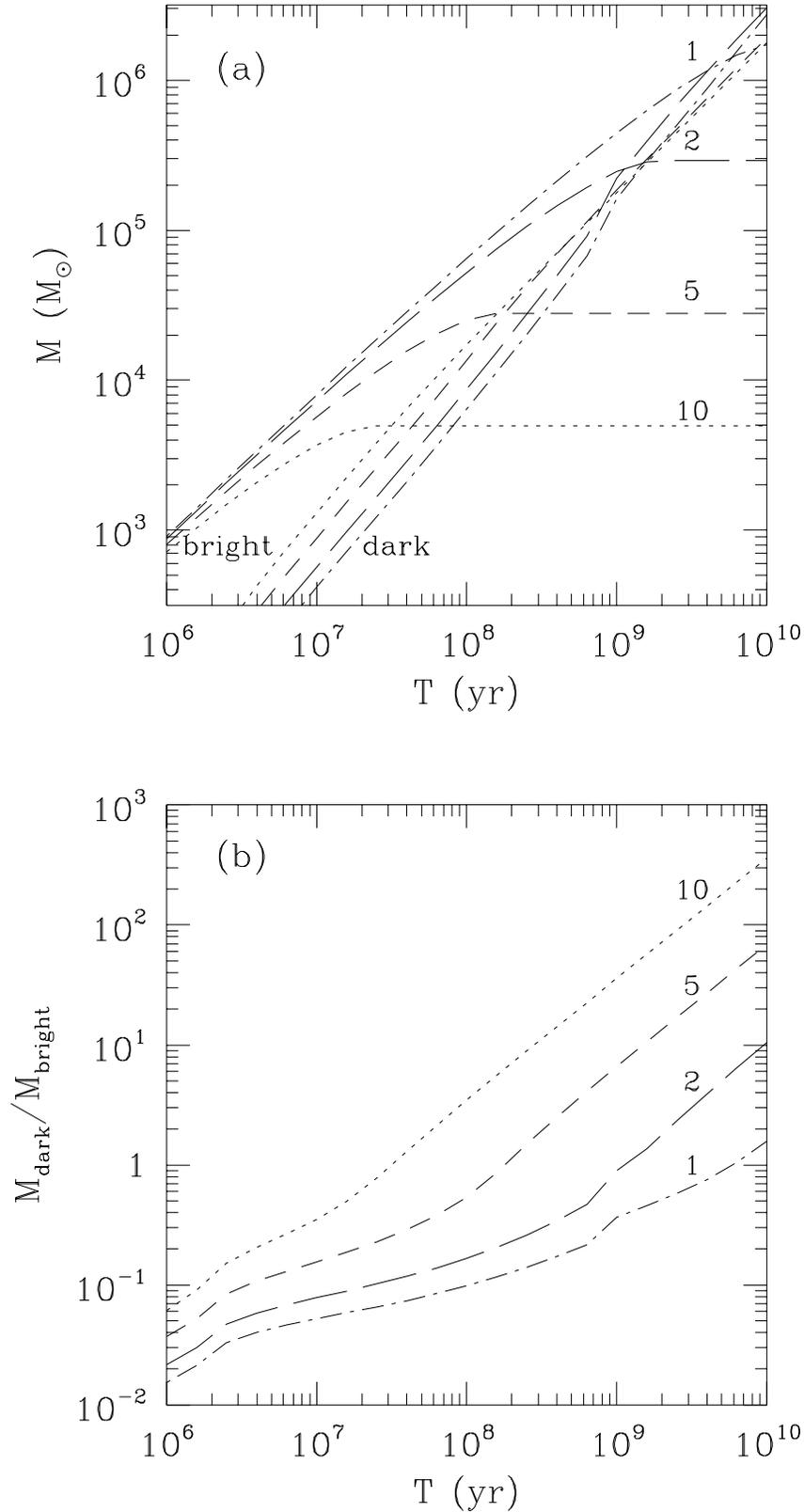

Fig. 10.— Masses of luminous stars and dark remnants resulting from continuous star formation as functions of the stellar population age. The upper panel (a) shows the total bright and dark star masses for Salpeter initial mass functions with an upper mass cut-off of 120 $M_\odot$ and lower mass cut-offs of 10 $M_\odot$ (*dotted lines*), 5 $M_\odot$ (*short dashed lines*), 2 $M_\odot$ (*long dashed lines*), and 1 $M_\odot$ (*dot-dashed lines*). The lower panel (b) shows the ratio of dark-to-bright mass for the same lower mass cut-offs.

### 3.2. The Distribution of Hot Stars

The preceding discussion and analysis have been based on the distribution and velocities of the "cool" stars. The compact surface brightness distribution of the "hot" stars differs from that of the "cool" stars (Allen 1994), and so is also of interest. We now show that the compact "hot" star surface brightness distribution is consistent with our inferred total mass distribution and the He I star velocities measured by Krabbe et al. (1995) and Haller et al. (1996), if the He I star velocity dispersion gradient is flatter than that of the "cool" stars.

Our derived total mass distributions (Figs. 5 and 7) are approximately linear over the inner $\sim 0.5$ pc (12″), so that we take $M(r) = \alpha r$ in this region with $\alpha = 1.1 \times 10^7 \, M_\odot \, \text{pc}^{-1}$ and $8.2 \times 10^6 \, M_\odot \, \text{pc}^{-1}$ for the mass profiles based on the first and second velocity dispersion fits, respectively. The density, $\rho_\text{h}$, of hot stars is therefore given by

$$\frac{1}{\rho_\text{h}} \frac{d}{dr} \left( \rho_\text{h} \sigma_\text{r}^2 \right) = -\frac{G\,M(r)}{r^2} = -\frac{G\,\alpha}{r} \qquad (11)$$

where, as before, isotropy is assumed. With an isothermal distribution ($\sigma_\text{r}$ = constant), it is straightforward to show that $\rho_\text{h} \propto r^{-\beta}$ where

$$\beta = \frac{G\,\alpha}{\sigma_\text{r}^2} = 4.3 \left( \frac{\alpha}{10^7 \, M_\odot \, \text{pc}^{-1}} \right) \left( \frac{\sigma_\text{r}}{100 \, \text{km s}^{-1}} \right)^{-2}. \qquad (12)$$

Krabbe et al. (1995) quote a value of 210 km s$^{-1}$ for the velocity dispersion of 12 He I stars within 0.22 pc (5.35″) of Sgr A*. However, this value is strongly influenced by one outlier with $V_\text{LSR} = +410 \pm 60$ km s$^{-1}$. Omitting this star gives a statistically corrected (see Da Costa et al. (1977)) velocity dispersion of $160 \pm 34$ km s$^{-1}$. Another high velocity star was given very low weight by Krabbe et al., so it does not influence the dispersion significantly. Haller et al. (1996) give velocities for 8 He I stars within 6″ of Sgr A* which also include one significant outlier. Discarding this star gives a statistically corrected velocity dispersion of $84 \pm 24$ km s$^{-1}$, assuming that their individual velocities have errors of $\sim 50$ km s$^{-1}$. The remaining 7 He I stars from Krabbe et al. (1995) lie between 0.26 pc (6.4″) and 0.48 pc (11.7″) from Sgr A* and have a statistically corrected velocity dispersion of $149 \pm 40$ km s$^{-1}$. We adopt the latter value as the velocity dispersion of the "hot" stars in the range 0.2 pc (5″) $\leq r_\text{p} \leq$ 0.5 pc (12″) where the linear mass profile appears to be valid. For this velocity dispersion and the two values of $\alpha$ derived from our two enclosed mass profiles, the "hot" star density distribution is predicted to have power-law indices of $\beta = 2.1 \pm 1.1$ and $1.6 \pm 0.9$, respectively. Although the velocity dispersion error leads to significant uncertainties in our estimates of $\beta$, the values we derive are both consistent with the actual value of 2.3 obtained by deprojecting the observed "hot" star distribution.

Why is the profile of the "hot" star distribution steeper than that of the "cool" stars when the projected velocity dispersions are similar at 0.2 pc? In our simple model, the answer to this lies in our assumption of isothermality. Consider the Jeans equation (rotation neglected) in the form:

$$M(r) = \frac{\sigma_r^2 r}{G} \left[ -\frac{d\ln\rho}{d\ln r} - \frac{d\ln\sigma_r^2}{d\ln r} \right] \qquad (13)$$

For the "cool" star population, both the density gradient and the velocity dispersion gradient contribute to the inferred mass. If the "hot" star population is isothermal, the velocity dispersion gradient is absent so a steeper density profile is required for the same mass distribution. Velocity dispersions based on the He I star data discussed above are too uncertain to define a meaningful velocity dispersion gradient. It therefore remains to be seen whether the "hot" star velocity dispersion profile is significantly flatter than that of the "cool" stars. Nevertheless, our isothermal model for the "hot" star density is at least consistent with the present data.

Krabbe et al. (1995) have used virial mass estimates based on the kinematics of He I stars to infer a mass of $\sim (4 \pm 1.6) \times 10^6 \ M_\odot$ within 0.14 pc of the Galactic center. This result conflicts with our inferred mass of $1.2$–$1.7 \times 10^6 \ M_\odot$ within 0.2 pc (subject to the caveat discussed earlier regarding the presence of a central minimum in the "cool" star distribution). The discrepancy principally arises from the high He I star velocity dispersion adopted by Krabbe et al. (1995), with a minor contribution from their inappropriate use of the virial theorem which we discuss in an Appendix. The combined effect of the lower He I velocity dispersion within 0.22 pc ($5.35''$) discussed above and the virial correction is to reduce the mass within 0.14 pc inferred from the He I stars to $0.77 \times (160/210)^2 \times (4 \pm 1.6) \times 10^6 \ M_\odot = (1.8 \pm 0.7) \times 10^6 \ M_\odot$. This value is in satisfactory agreement with the range of values derived from our "cool" star analysis.

## 4. CONCLUSIONS

We have presented a reanalysis of the dynamics of the Galactic center star cluster incorporating significant refinements over the treatment of MSBH, SMBH, and Krabbe et al. (1995). From this we have derived a radial mass profile for the inner 4 pc, and extended the analysis by deriving the radial dependence of $M/L_K$, the $K$ band mass-to-light ratio. Within the limitations of the assumptions of our dynamical model, we have found that $M/L_K$ may increase with decreasing radius within $\sim 0.8$ pc from the Galactic center. Such a variation of $M/L_K$ cannot be explained by the existence of a central point mass and, if correct, indicates that the nature of the stellar population changes in the central region.

In order to explain this varying mass-to-light ratio, we have suggested that the central star cluster may have experienced successive starbursts that have favored the formation of massive stars, possibly similar to but smaller than the events seen in starburst galaxies. In such a scenario, the evolved remnants of previous generations of massive stars would now populate the central region of the Galaxy and dominate the mass within $\sim 1$ pc of the Galactic center.

We have presented one possible interpretation of the stellar dynamics of the Galactic center. However, uncertainties in the published data and the possibility of a central minimum in the "cool" star distribution prevent us from clearly distinguishing between various other models. Equally acceptable fits to the available data based on a central point mass are certainly possible (e.g., Kent 1992). A conclusive decision on the existence or otherwise of a massive black hole requires velocity dispersion data significantly inside the current 0.2 pc projected radius limit. Uncertainties in the spatial distribution of CO absorbing stars and possibly in the atmospheric dynamics of the He I emission-line stars make proper motion measurements the most reliable means of determining velocity dispersions. If Sgr A$^*$ is a $10^6\ M_\odot$ black hole, the 10 objects detected by Eckart et al. (1995) within a projected distance of 0.02 pc (0.5″) of Sgr A$^*$ should have a velocity dispersion $\geq 270$ km s$^{-1}$ leading to detectable proper motions with HST on a timescale of a few years. On the other hand if the central dark mass is distributed, the proper motions of these objects would be up to a factor of $\sim 2$ lower.

As we have demonstrated in § 2.4, uncertainties in our fits to the available velocity dispersion data lead to significantly different mass profiles which affect conclusions about the nature of the stellar population. Consequently, it is important to determine the velocity dispersion profile more precisely, not only at small radii, but also at radii around 1 pc. We also emphasize the value of determining the inferred mass as a function of luminosity since this simple diagnostic has the potential to be extremely revealing. In our analysis, for example, it has laid open the possibility of a significantly varying mass-to-light ratio in the Galactic center.

We acknowledge a number of enlightening discussions on the Galactic center with the late David Allen who also provided us with his and Michael Burton's photometric data used in our analysis. David Allen made an enormous contribution to the study of the Galactic center during an extremely productive and energetic scientific career. We also thank Agris Kalnajs for providing a deprojection subroutine, Gary Da Costa for his critical reading of an earlier version of this paper, Joe Silk for suggesting a truncation of the initial mass function, and our referee Alfred Krabbe for helpful and constructive comments.

## A. Virial Mass Estimates

The virial theorem is usually applied to isolated (or nearly isolated) systems. However, the hot stars in the Galactic center cluster are not an isolated self-gravitating system, so a modified form of the virial theorem should be used to estimate enclosed mass. When the system is not isolated, the development of the virial equation (see Binney & Tremaine (1987), p211 ff) requires the addition of a surface integral so that it reads:

$$\int_V \rho \sigma_{ii} \, d^3x - \int_S \rho \, \sigma_{ij} \, x_j n_i \, dS + W = 0 \qquad (A1)$$

where $\sigma_{ij} = <v'_i v'_j>$ is the velocity dispersion tensor, $v'_i$ are the stellar velocities, $W$ is the gravitational potential energy, and $S$ is the surface, with unit normal $n_i$, bounding the volume $V$. The surface integral represents the effect of a non-isolated system. The physical reason for its existence is that stars currently within the volume $V$ are not bound by the mass within $V$. Their orbits take them outside so that the gravitating mass implied by the kinetic energy within $V$ is reduced. Similar considerations also apply to the use of the Bahcall-Tremaine (1981) mass estimator.

In order to assess the effect of the surface integral, we assume spherical symmetry, take $S$ to be a spherical surface with radius $R$ and, as throughout this paper, assume that the velocity dispersion is isotropic. The above virial equation (A1) becomes:

$$4\pi \int_0^R \rho \, \sigma^2 \, r^2 \, dr - \frac{4\pi}{3} \rho(R) \, \sigma^2(R) R^3 = 4\pi G \int_0^R M(r) \, \rho r \, dr. \qquad (A2)$$

This equation may also be derived directly from the Jeans equation (1) (with $v_{\rm rot} = 0$) by multiplying by $r^3$ and integrating. However, we have taken the virial theorem route here in order to clarify the relationship between the use of stellar hydrodynamics and the virial theorem. The virial estimate of mass is reduced by the second term on the left, and in order to estimate its effect we evaluate the fraction

$$f = \frac{\frac{4\pi}{3} \rho(R) \sigma^2(R) R^3}{4\pi \int_0^R \rho \sigma^2 r^2 \, dr}. \qquad (A3)$$

Taking the velocity dispersion to be constant, as we assume for the Galactic center "hot" star population, and the stellar density $\rho_* \propto r^{-\beta}$ gives $f = (3-\beta)/3$, independent of radius. For the Galactic center "hot" stars, $\beta = 2.3$ giving $f = 0.23$ so that the virial estimates of Krabbe et al. (1995) should be corrected by the factor $1 - f = 0.77$.

## REFERENCES

Allen C. W., 1973, Astrophysical Quantities, 3rd edn., (Athlone Press, University of London).